\documentstyle[12pt]{article}
\def\lsim{\lower0.6ex\vbox{\hbox{$ \buildrel{\textstyle <}\over{\sim}\ $}}}
\def\rsim{\lower0.6ex\vbox{\hbox{$ \buildrel{\textstyle >}\over{\sim}\ $}}}
\def\beq{\begin{equation}}
\def\eeq{\end{equation}}
\def\beginapjbib{\begingroup \section*{\large \bf References}
         \parskip=.5ex plus 1.0pt
         \def\bibitem{\par \noindent \hangindent\parindent
                \hangafter=1}}
\def\endapjbib{\par \endgroup}
\begin{document}
\begin{flushright}
UMN-TH-1230/94\\
OSU-TA-6/94  \\
April 1994
\end{flushright}
\vskip 0.75in

\begin{center}

{\Large{\bf   On the Abundance of Primordial Helium}}

\vskip 0.5in
{  Keith~A.~Olive$^1$  \\ and \\ Gary~Steigman$^{2,3,4} $}

\vskip 0.25in
{\it
$^1${School of Physics and Astronomy,
University of Minnesota, Minneapolis, MN 55455}\\

$^2${Department of Physics,
The Ohio State University, Columbus, OH 43210}\\

$^3${Department of Astronomy, The Ohio State University, Columbus, OH 43210}\\

$^4${Institute of Astronomy, University of Cambridge, Cambridge CB3 0HA,
England}\\
}

\newpage

{\bf Abstract}
\end{center}

We have used recent observations of helium-4, nitrogen and oxygen
from some four dozen, low metallicity, extra-galactic HII regions to
define mean $N$ versus $O$, $^4He$ versus $N$ and $^4He$ versus
$O$ relations which are extrapolated to zero metallicity to determine
the primordial $^4He$ mass fraction $Y_P$.  The data and various subsets
of the data, selected on the basis of nitrogen and oxygen, are all
consistent with $Y_P = 0.232 \pm 0.003$.  For the 2$\sigma$
(statistical) upper bound we find $Y_P^{2\sigma} \le 0.238$.
Estimating a 2\% systematic uncertainty $(\sigma _{syst} = \pm 0.005)$
leads to a maximum upper bound to the primordial helium mass
fraction:  $Y_P^{MAX} = Y_P^{2\sigma} + \sigma_{syst} \le 0.243$.
We compare these upper bounds to $Y_P$ with recent calculations of
the predicted yield from big bang nucleosynthesis to derive upper
bounds to the nucleon-to-photon ratio $\eta$ ($\eta_{10} \equiv
10^{10}\eta$) and the number of equivalent light (\lsim 10 MeV)
neutrino species.  For $Y_P \le 0.238$ ($0.243$), we find $\eta_{10}
\le 2.5 (3.9)$ and $N_\nu \leq 2.7 (3.1)$.  If indeed $Y_P \le
0.238$, then BBN predicts enhanced production of deuterium and
helium-3 which may be in conflict with the primordial abundances
inferred from model dependent (chemical evolution) extrapolations
of solar system and interstellar observations.  Better chemical
evolution models and more data - especially $D$-absorption in the
QSO Ly-$\alpha$ clouds - will be crucial to resolve this potential crisis for
BBN.  The larger upper bound, $Y_P \leq 0.243$ is completely
consistent with BBN which, now, bounds the universal density of
nucleons (for Hubble parameter $40 \le H_o \le 100~{\rm km s}^{-1}
{\rm Mpc}^{-1}$ and cosmic background radiation temperature T = $2.726 \pm
0.010$) to lie in the range
 $0.01 \le \Omega_{BBN} \le 0.09$ (for $H_o =
50h_{50} {\rm km s}^{-1}
{\rm Mpc}^{-1}$, $0.04 \le \Omega_{BBN} h^2_{50} \le 0.06$).

\newpage

\noindent
\section{Introduction}

After hydrogen, helium-4 is the next most abundant nuclide in the Universe.  As
a result,
 the
primordial abundance of $^4He$, synthesized in the first $\sim$ 20 minutes of
the
 evolution of the Universe, has assumed a crucial role in testing the standard
hot big bang model of cosmology and, in placing constraints on particle physics
beyond the standard (Glashow-Weinberg-Salam) model.  As a consequence of its
high abundance, $^4He$ can be observed throughout the Universe; this is in
contrast to the
other light elements ($D$, $^3He$ and $^7Li$) produced during   big bang
nucleosynthesis
(BBN) which are, to date, only observed in the Galaxy.  In addition, the $^4He$
abundance may be
determined to much higher accuracy ($\sim$ few percent) than is the case for
the
other nuclides synthesized in BBN.

In the context of standard BBN, the predicted primordial abundance of $^4He$ is
large (the mass fraction, denoted by $Y_{BBN}$, is $ \sim 0.24$; the
ratio by number to hydrogen,
$y{^{BBN} _4} \sim 0.08$) and insensitive (logarithmically dependent) to the
one free parameter -- the nucleon-to-photon ratio $\eta~(\eta \equiv
n_N/n_{\gamma};
 \eta_{10} \equiv 10^{10} \eta$).  For a review and further references, see
Boesgaard \& Steigman
(1985); for  recent status reports see Walker et al. (1991; WSSOK) and Reeves
(1993).
$Y_{BBN}$ is sensitive to the expansion rate of the early Universe which, at
the epoch of
BBN, provides a measure of the total energy density (Shvartsman 1969).
Therefore,
the comparison between the accurately predicted $Y_{BBN}$ with $Y_P$ derived
from
accurate observational data (hereafter we will distinguish the primordial
abundance
inferred from observations, $Y_P$, from the primordial abundance predicted by
BBN,
$Y_{BBN}$), is the keystone of the consistency tests of the standard model of
cosmology and, provides constraints on new physics beyond the standard model of
particle physics (Steigman,
Schramm \& Gunn 1977).   To take full advantage of this test of cosmology and
approach
to high energy physics clearly requires accurate values of $Y_P$.

The derivation of $Y_P$ from astronomical observations is complicated by the
fact that
in the course of their evolution stars burn hydrogen to helium and, when they
die,
they return this processed material to the interstellar medium (ISM) polluting
the primordial
$^4He$.  To minimize the contribution from stellar-produced $^4He$, we
concentrate on measurements of the helium abundance in those regions whose low
heavy
element abundances suggest the least contamination from stellar and galactic
chemical evolution.  This has led virtually all investigators to the low
metallicity, extragalactic
HII regions (Searle \& Sargent 1971; Kunth \& Sargent 1983; Lequeux et al.
1979;
Pagel et al. 1992; Skillman \& Kennicutt 1993; Skillman et al. 1994a,b).
  Even for these data, from regions
whose metallicities range
down to 1/40 of solar, a correction for newly synthesized $^4He$ must be made.
The
standard approach has been that of Peimbert \& Torres-Peimbert (1974), to
correlate $Y_{OBS}$ with metallicity and, extrapolate to zero metallicity to
infer
$Y_P$.  Since the heavy element mass fraction, $Z$, is not observed,
 the observed
abundances of oxygen and/or nitrogen have usually served as surrogates for $Z$.

Another reason for concentrating on the lowest metallicity extragalactic HII
regions is
that the $^4He$ abundance is derived from the recombination lines of singly
and doubly ionized $^4He$; neutral $^4He$ is unobserved.  If the HII and HeII
zones do not
coincide, the neglect of HeI will introduce errors into $Y_{OBS}$.  Model HII
region
calculations show that for the highest excitation regions, ionized by the
hottest stars, the
HII and HeII zones do coincide (to $\sim$ 1\%; Skillman et al. 1994a).
 The more metal-poor
stars
are hotter and, therefore, by restricting attention to the most metal-poor HII
regions,
this systematic correction may be minimized.

Recently, Pagel et al. (1992; PSTE) have assembled a large data set (several
dozen) of
extragalactic HII regions observed/analyzed in a homogenous fashion.  The $Y$
vs. $O/H$ and/or $N/H$ correlations in this
 data set have been analyzed (Pagel et
al. 1992;
Olive, Steigman \&Walker 1991 (OSW); Fuller, Boyd \& Kalen 1991; Pagel \&
Kazulaskis 1992;
Mathews, Boyd \& Fuller 1993; Pagel 1993) to derive $Y_P$.  Virtually all
analyses agree that $ 0.22~ \lsim Y_P ~\lsim 0.24$.  The problems -- and
disagreements -- arise in the quest for the 3rd significant figure in $Y_P$.
For example,
is $Y{_P ^{MAX}} = 0.240$ or  0.243 or  0.237 (OSW)?  To approach $Y_P$ at the
1--2 \% level
requires great care with the statistics of
 the $Y$ vs. $O/H$ or $N/H$ fits, great care
in selecting the data sets and, an understanding of possible systematic
effects.  For
example, recently the ``standard" He emissivities of Brockelhurst (1972) have
been
challenged by Smits (1991).  Though these latter emissivities
 were found to be in error (Smits,
private communication to Skillman), such a correction could in principle
increase
$Y_{OBS}$ systematically
by up to 3\% ($\Delta Y \approx 0.007$)(Skillman \& Kennicutt 1993).

The analysis we present here was stimulated by the desire to determine $Y_P$
(or, at least,
$Y{_P ^{MAX}}$) to $\sim 2\%$ accuracy (or better) for comparison to $Y_{BBN}$
in tests of cosmology
and particle physics.  It was encouraged by the valuable addition of 11 new,
very metal-poor
HII regions (Skillman et al. 1994a,b). Skillman et al. (1994b) have
graciously provided us with preliminary results of their data analysis.

In the next section we analyze the $N$ vs. $O$ correlations in the PSTE and
Skillman et al. (1994a,b) data sets with the goal of resolving the questions of
secondary
versus primary
nitrogen (Fuller et al. 1991; Pagel \& Kazulaskis 1992; Mathews et al. 1993),
of
possible Wolf-Rayet contamination (PSTE) and, of how best to choose a
sufficiently homogeneous metal-poor data set to use for exploring $Y_P$.  Then,
using
the subset(s) of the PSTE and (preliminary)
Skillman et al. (1994a,b) data we have
identified from the nitrogen and oxygen data, we study the $Y$ vs. $O/H$ and
$N/H$
correlations to infer
$Y_P$.  Since, for some tests of  particle physics and cosmology we may wish to
compare
$Y{_P ^{MAX}}$ with $Y{^{MIN} _{BBN}}$, we comment on the uncertainty in $Y{_P
^{MAX}}$.
Armed with $Y_P$ from our statistical analysis, we next compare to $Y_{BBN}$
from the
latest BBN calculations (Kernan 1993; Kernan, Steigman \& Walker 1994) to
derive
constraints on the consistency of BBN, on the nucleon abundance ($\eta$) and on
particle physics beyond the standard model ($N_{\nu}$).  Finally, we
summarize
our conclusions and their implications for cosmology, for particle physics, and
for further
astronomical observations.

\section{Classifying Metal-Poor HII Regions}

In Figure 1 we show the nitrogen and oxygen abundances observed for all the 49
HII Regions
 in the  PSTE  and Skillman et al. (1994b) data sets.  Although low
metallicity extragalactic HII regions are surely not a homogeneous set, we do
want -- to the extent possible -- to identify a nearly primordial, relatively
unpolluted (by the
products of stellar evolution) subset.  From Figure 1 it is clear that if we
focus
on those HII regions with $N/H \leq 1.0 \times 10^{-5}~ {\rm and}~ O/H \leq 1.5
\times 10^{-4}$,
the 8 HII regions we discard have significantly higher nitrogen and/or oxygen
abundances
than the 41 HII regions we retain.  Our ``first cut"
 metal-poor data set spans one order of
magnitude
in oxygen abundance ($15~ \lsim 10^6 (O/H)~ \lsim 150$) and a factor of $\sim
25$ in nitrogen
abundance ($4 ~\lsim 10^7 (N/H)~ \lsim 100$).  Although the iron abundance in
these
HII regions is unknown, we may estimate $[Fe/H]$ for our metal-poor data set
using oxygen
and/or nitrogen as surrogates for iron.  The solar oxygen abundance (Grevesse
\& Anders
1989) is $[O]_{\odot} \equiv 12 + \log (O/H)_{\odot} = 8.93$ so that for
$[O/H]$
$\equiv
[O] -[O]_{\odot}$ we have, $-1.75 ~\lsim [O/H] \leq -0.75$.  From studies of
metal-poor stars
 it has been noted that oxygen (and, perhaps, other $\alpha$-nuclei as well) is
enhanced
with respect to iron:  $ [O/Fe] \approx $0.5 (Sneden, Lambert \& Whitaker
1979; Barbuy \& Erdelyi-Mendes 1988; Wheeler, Sneden \& Truran 1989) so that we
may
infer for our
metal-poor HII regions that $-2.25 ~\lsim [Fe/H] ~\lsim -1.25$.  If, instead,
we compare
to nitrogen: $-2.45 ~\lsim [N/H]~ \lsim -1.05$ (with a similar estimate for
$[Fe/H]$ since
$[N/Fe] \approx 0$ for metal-poor stars).  Thus, we are dealing with a sample
whose
contamination (compared, e.g., to the Galaxy) is relatively small.

Later, to probe the robustness of our statistical results, we will also
make a ``second cut" and consider a
very metal-poor subset; here we will choose the 21 HII regions with
 $O/H \leq 8  \times 10^{-5}$.  Again,
there is a ``gap" in oxygen abundance between the 21 regions we keep and the 20
we
discard.  This very-metal-poor set has modest dynamical range with
 $ 15 ~\lsim 10^6 ~(O/H)~ \lsim 80  ~{\rm and}~ 4 ~\lsim 10^7 (N/H) ~\lsim
{}~40$.

Although a study of the nitrogen versus oxygen relation for metal-poor HII
regions is of
great intrinsic interest for the study of chemical evolution, it must be
emphasized
that such small regions are likely dominated by local -- in space and in time
--
processes.  Different regions may be ``caught" at different evolutionary epochs
(e.g.,
just before or just after a starburst).  Thus, a study of the $N/H$ versus
$O/H$
relation for extragalactic HII regions need not shed much light on the chemical
evolution of our -- or, any other individual -- galaxy.  Chemical evolution
models (e.g., Mathews, Boyd \& Fuller (1993)) may provide a guide which this
data set
ignores.  Here, we are simply hoping to exploit the low oxygen and nitrogen
abundances as an aid in extrapolating the helium abundance to its
uncontaminated--
primordial -- value.  In so doing, we implicitly presume that for low
metallicity regions there exist
mean relations among $N$ and $O$, $^4He$   and $O$,
$^4He$  and    $N$.  However,
we do have to concern ourselves with those local processes which may have
introduced excess dispersion
in the $N$ vs. $O$, $He$ vs. $O$ and $He$ vs. $N$
 relations we infer from the data.

Indeed, Pagel, Terlevich and Melnick (PTM, 1986) noted that some HII regions
with observed
Wolf-Rayet spectral features often had larger abundances of both helium and
nitrogen compared
to other regions with the same oxygen abundance.  PTM suggested that such
regions may have
temporary excesses of $He$ and $N$ due to pollution
 from stellar winds containing
the
products of hydrogen burning.  PSTE identify those HII regions in their set
which have
detected (D) and/or possible (P) WR features in their spectra and they
distinguish them from
those ``clean" (C) regions lacking such spectral features.  PSTE argue that
analysis
of the data should be restricted to those objects (C or, perhaps, C + P) for
which there
is no evidence that the pollution effect is present.  However A. Maeder, in the
discussion of Pagel's paper (1991) at IAU Symposium No.149, notes that helium
will be
ejected also before the stars reach the WR phase and, that since the WR phase
is
very short-lived, the absence of WR spectral features is not evidence of
absence of
WR pollution.  Thus, in a statistical sense, it may not be justified to exclude
HII regions
from the analysis simply on the basis of the presence (D) or absence (C) of
WR features.  Therefore, one of our first goals is to explore whether, based on
the
nitrogen and oxygen data alone, there are statistically significant differences
between
the  C, P and D data sets (note that all the HII regions in the Skillman et al.
sample are ``C").

In our first approach to the $N$ vs. $O$ relation,
 we have fit the data from all 41
regions
of our
metal-poor data set to a power law of the form $N/H = A(O/H)^{\alpha}$; we have
also fit the C, P and D subsets (with, respectively, 22, 7 and 12 regions).  In
Table 1
we display the correlation coefficients (r) and the reduced chi-squared
$({\chi}^2/dof)$
along with A and $\alpha$ for our fits.
\vskip .5in
\begin{table}[h]
\centerline {\sc{\underline{Table 1:} Power Law fits to $N$ vs. $O$}}

\vspace {0.1in}
\begin{center}
\begin{tabular}{|cccccc|}                     \hline \hline
Set &  \# Regions & $r$ & ${\chi}^2/dof$ &  $A$  &  $\alpha$  \\ \hline

All & 41 & 0.91 & 2.1 & 0.76 & $ 1.31 \pm 0.07$ \\
C & 22 & 0.94 & 1.3 & 0.20 & $ 1.18 \pm 0.08$ \\
P & 7 & 0.91 & 3.2 & 828 & $2.06 \pm 0.57$ \\
D & 12 & 0.70 & 2.1 & 15.5 & $1.65 \pm 0.33$\\
P+D & 19 & 0.81 & 2.5 & 98.0 & $1.84 \pm 0.30$ \\\hline
\end{tabular}
\end{center}
\end{table}

Although the P and D sets do seem to differ from the C set, it is noteworthy
that the fit for the D
set (WR features observed) is as close to that for
the C set as it is to the fit for the P set.  Indeed,
although the
numbers are small and the uncertainties  large, it is the P set (possible WR
features)
which seems anomalous.  In any case, it seems difficult to argue that the C and
D sets
 differ
statistically. It is clear from the reduced
$\chi^2$s that there is more dispersion about the mean fits
for the P and D data sets - especially for the P set - than
for the C set. Thus, although it is likely that the PTM effect is present in
our
sample
-- as well as other effects we discuss below -- on the basis of the
observed nitrogen
and the
oxygen abundances alone, it is not possible to
 identify individual ``contaminated" HII
regions.
Therefore, initially, we will continue to use all 41 HII regions in our
subsequent
analyses.

The power $\alpha = 1.3 \pm 0.1$ of our fit to the $N$ vs. $O$
 data supports neither
a
``primary" (linear $N$ vs. $O$ relation) nor a ``secondary"
(quadratic $N$ vs. $O$
relation) origin for nitrogen in the metal-poor HII regions.
However, that for the C-set alone, $\alpha_C = 1.2 \pm 0.1$ is
marginally consistent with a purely linear relation. Remember, though,
that
this heterogeneous sample is not likely to track the galactic evolution of
the nitrogen and oxygen abundances.  Indeed, the poor chi-squareds of our fits
are, at least
in part, due to the dispersion in the data.  However, to further explore the
``primary" versus
``secondary" nature of the $N$ vs. $O$ relation
(for our purposes here, ``primary"
and
``secondary" should be replaced by ``linear" and ``quadratic" respectively), we
have evaluated the $N/O$ ratio for each of the 41 HII regions in our
first cut sample and we have fit the data
to $N/O = a + b(O/H)$.  The data is displayed
in Figure 2
along with our best fit
\beq
10^2 (N/O) = 2.5 \pm 0.3 + (1.4 \pm 0.4) \times 10^4 (O/H).
\label{lin}
\eeq
\noindent This data ($N/O$ vs. $O/H$) is not strongly
 correlated ($r = 0.31$) and the
${\chi}^2/dof = 2.3$ is not an improvement over our previous power-law fit
(also
a 2-parameter fit).  For our first cut metal-poor sample, the linear term (in
$N$ vs. $O$)
dominates over the quadratic one; only for $10^4 (O/H)~ \rsim 1.8$ does the
quadratic
term exceed the linear.  Here, we are in agreement with Pagel and Kazlauskas
(1992)
who also conclude that ``primary"  nitrogen dominates at low metallicity.  We
both
disagree with the claim of Mathews, Boyd and Fuller (1993) that virtually all
the
nitrogen is secondary.  This claim is repeated by Balbes, Boyd and Mathews
(1993) who add
the caveat that the primary contribution may dominate at times or metallicities
for
which there is a paucity of data.  This latter point is no longer true given
the recent,
very low metallicity data of Skillman et al. (1994b).

Given the weak $N/O$ vs. $O/H$ correlation,
 we have also evaluated the (weighted)
mean
$N/O$ ratio for our data set,
\beq
10^2 \langle N /O \rangle = 3.4 \pm 0.2
\label{ave}
\eeq
where the error in (\ref{ave}) is the error in the mean and does not represent
the scatter in the data.
For this fit the ${\chi}^2/dof = 3.0$; by the F-test (Bevington 1969) this is
not as good a fit, at greater than the 99.9\% confidence level, than
either of the two-parameter
 power-law or linear/quadratic fits in (\ref{lin}).
Indeed, the poor chi-squareds of our fits suggest that there may indeed be real
dispersion about a mean $N$ vs. $O$ relation.
 If so, this may well bias our $He$ vs.
$N$ or $O$ fits.
And, so, we examine this issue further in the following.

We have already mentioned the PTM suggestion that HII regions with WR spectral
features may be contaminated with excess $N$ and $He$ (relative to their $O$
abundance).  There are
other sources of dispersion for extragalactic HII regions.  For example, in a
region where
there has been a recent starburst, the HII region may have been contaminated by
the
products of the evolution of the most massive stars.  Such regions would have
excess
$O$ (relative to their $N$ abundance) and slightly enhanced $^4He$.  For the
observed $N/H$
ratio, $N/O$ will be {\it low} for such regions (for the observed $O/H$ ratio,
$N/O$ will also be {\it low}).  However, at the observed $N/H$, there will be
``extra" $^4He$ causing an
{\it upward} dispersion from a mean $Y$ vs. $N/H$ relation.
  However, at the observed
(``excess")
$O/H$, there isn't the ``normal" contribution to $^4He$ (from the lower mass
stars) so that there will be a {\it downward}
dispersion from a mean $Y$ vs. $O/H$
relation.  This is a counter-example to the claim of Campbell (1992) that the
dispersion in $Y$ vs. $O/H$ will be, ``one-sided, i.e. upward from a minimum
value
of $He/H$ at each $O/H$" and argues
against her proposal that the primordial helium
abundance can only be reliably determined by fitting to the {\it lower}
envelope
of the $Y$ vs. $O$ relation.

As another example, suppose that stellar winds and supernovae combine to blow a
superbubble (De Young \& Gallagher 1990)  in the HII region.  Such regions may
have lost some of
their
oxygen (and, the accompanying helium) but retained the products of the
longer-lived stars (e.g., $N$ and $^4He$).
For such regions the observed $N/O$ ratio will be
{\it high}
and there will be an {\it upward} dispersion in the $Y$ vs. $O/H$ relation
(more
$^4He$ from low mass
stars relative to the observed oxygen abundance) but, a {\it downward}
dispersion in the $Y$ vs. $N/H$
relation (some $^4He$ has been lost from the system).

Finally, we return to the PTM effect.  If WR
activity has contaminated the HII region with ``extra" $N$ and $^4He$ (for its
observed
oxygen abundance) then $N/O$ will be {\it high} and there will be an {\it
upward}
fluctuation
in the $Y$ vs. $O/H$ relation and a {\it downward} fluctuation in the $Y$ vs.
$N/H$
relation
(since, for the observed $N/H$, the $O/H$ is low,  so too will be
the $^4He$ contribution corresponding to $O/H$).

The effects outlined above show that individual HII regions may experience
either
upward or downward excursions in $N/O$.  Thus, we have searched our data set to
identify
such ``outliers".
To search for
discrepant HII regions we have placed two sigma
contours around each data point in Figure 2 and asked if any of them
do not cross the linear fit in eq. (\ref{lin}).
In this manner, we have identified
seven regions which we consider to be ``outliers": T 1304-38 (4.9$\sigma$,P),
N 4861 (2.7$\sigma$, D),
II ZW 40 (2.7$\sigma$,D),
TOL 65 (2.7$\sigma$,P), TOL 35 (2.4$\sigma$,D), CS 0341-40 (2.1$\sigma$,P),
and  SBS0335 (2.0$\sigma$,C),. They are listed in the order of most
to least discrepant
(with the discrepancy given in terms of the quoted errors, and the
Wolf-Rayet characteristic of the region).
Note that six of these seven outliers are P or D. One of the two outliers
which have relatively
low $O/H$ ($10^6 O/H < 60$)  has low $N/O$ while the other has high
$N/O$. Of the five regions with higher $O/H$, three have low $N/O$
and two have high $N/O$. Thus, as may be seen in Figure 3, where the
outliers are identified,
there is no general trend
in these discrepant regions.

With these outliers eliminated, we have
refit the $N/O$ vs. $O/H$ relation as well as recalculated $\langle N/O
\rangle$ for
the remaining 34 HII regions.
\beq
10^2 \langle N/O \rangle_{34} = 3.4 \pm 0.2,
\eeq
\beq
10^2 (N/O)_{34} = 2.5 \pm 0.2 + (1.3 \pm 0.3) \times 10^4 (O/H).
\label{lin32}
\eeq
For the weighted mean the new ${\chi}^2 /dof ~{\rm is}~ 1.6$; for the linear
fit the correlation
coefficient is $ r = 0.4 ~{\rm and} ~    {\chi}^2 /dof = 0.96$.
Notice the marked improvement in the $\chi^2$.
 With respect to the latter fit (\ref{lin32})
there are no further outliers in the
remaining 34 HII regions. In figure 3, we show the same data (as in figure 2)
with the new fit (\ref{lin32}) and the outliers identified as filled circles.
We note that a power law fit to these 34 HII regions, with A = 0.44 and
${\alpha} = 1.26 \pm0.05$, also has an excellent reduced ${\chi}^2$
(${\chi}^2$/dof
= 0.93)

Statistically, the linear fit is preferred over the simple weighted mean
indicating a correlation between $N/O$ and $O/H$ and hence the presence
of {\em some} secondary nitrogen.  This is supported by our power law fit
where ${\alpha}$ differs from unity by some 5 sigma. We note that
 for this subset of our set of low
metallicity HII regions, the ``primary" component dominates; a secondary
component
would dominate only for $10^4 O/H > 1.9$, beyond the upper bound to the
oxygen abundances
for our data set.  Although there is no justification to extrapolate our fit
beyond $10^4 O/H= 1.5$, we note that for $[Fe/H] \approx [N/H]~\lsim -1,
[N/O]_{34}$ ranges from -0.4 to -0.7 which is not an unreasonable fit
 to the $[Fe/O]$ relation observed in halo stars (Sneden, Lambert \& Whitaker
1979;
Barbuy \& Erdely-Mendes 1988; Wheeler, Sneden, \& Truran 1989).
Indeed, (\ref{lin32}) only slightly overestimates the solar $N/O$ ratio
($10^2 (N/O)_{34} \approx 14$ vs. $10^2 (N/O)_{\odot} \approx 13$ for
$10^4 (O/H)_\odot \approx 8.5)$ and slightly underestimates the Orion $N/O$
ratio ($10^2 (N/O)_{34} \approx 7.8$ vs. $10^2 (N/O)_{Orion} \approx 11$ for
$10^4 (O/H)_{Orion} \approx 4.1$; Gies \& Lambert 1993 and Cunha \& Lambert
1993).

As emphasized at the outset, the goal is to identify a sufficiently large,
sufficiently
metal-poor sample so that the extrapolation to zero metallicity is minimal and
statistically
meaningful.  Our confirmation of the Pagel and Kazlauskas (1992) conclusion
that ``primary" nitrogen dominates for low O/H ($ \lsim 1.8 \times 10^{-4}$)
suggests
that we further consider a very low metallicity subset of our metal-poor HII
regions.
Half -- 21 -- of our 41 HII regions have $ 10^6 (O/H) ~\leq 80$; the
next
highest oxygen abundance is $10^6 (O/H) = 94 \pm 6$ (more than 2$\sigma$
higher).
For this subset we find
\beq
10^2 \langle N/O \rangle_{21} = 3.1 \pm 0.2,
\eeq
\beq
10^2 (N/O)_{21} = 2.1 \pm 0.4 + (2.5 \pm 1.0) \times 10^4 (O/H).
\label{lin20}
\eeq
For the weighted mean the ${\chi}^2/dof = 2.0$; for the linear fit the
correlation coefficient is $r = 0.37 ~{\rm and ~the} ~ {\chi}^2/dof = 1.5$.
For a power law fit we find A = 0.74, ${\alpha} = 1.30 \pm 0.12$, r = 0.89
and ${\chi}^2/dof = 1.4$.

In this reduced "second cut" set of 21 points we can repeat our previous
procedure to look for outliers
with respect to the fit (\ref{lin20}).
 Only the two outliers with
$10^6 O/H < 80$ already identified above (see fig.3) are found to be more
than
2$\sigma$ discrepant with the fit (\ref{lin20}).
 The resulting fit to the remaining 19 HII regions
is $10^2 (N/O)_{19} = 2.1 \pm 0.4 + (2.6 \pm 0.8) \times 10^4 (O/H)$ with a
$\chi^2/dof = 0.93$ and $r=0.49$. The corresponding power law fit has A =
1.4, ${\alpha} = 1.37 \pm 0.11$, r = 0.94 and $\chi^2/dof = 0.91$.

\section{Towards the Primordial Abundance of $^4He$}

The goal of our analysis is to use the $^4He$, $N$ and $O$ data from
the metal-poor extra galactic HII regions to define a mean $Y$ vs.
$N/H$ or $Y$ vs $O/H$ relation to be used to extrapolate to zero
metallicity to infer the primordial abundance of $^4He$, $Y_P$.  Since
the evolutionary history of higher metallicity HII regions may differ
from those more metal-poor, we have culled the 49 PSTE and
Skillman, et al. HII regions to a first cut set of 41 regions with
$10^6 O/H \leq 150$ and $10^7 N/H \leq 80$.  For this first cut set
we have explored the PTM and PTSE suggestions that HII regions
with observed WR features may have enhanced nitrogen (and,
possible $^4He$) relative to its oxygen abundance.  Although we find
no strong evidence supporting such a view, we did identify enhanced
dispersion about a mean (power law) relation for those regions with
detected (D) or possible (P) WR features.  So, in our $Y$ vs. $O/H$
and $Y$ vs. $N/H$ fits, we will consider the C set (22 regions) as well
as the full (41 regions) first cut set; for completeness we will also
calculate fits for all 49 PTSE and Skillman, et al. HII regions.

In our power law and linear plus quadratic $N$ vs. $O$ fits for our
first cut set we found relatively high reduced chi-squareds.
This stimulated us to search for ``outliers", regions whose $N/O$ ratio
was more than 2$\sigma$ discrepant (accounting, simultaneously, for the
uncertainties in $N/H$ and $O/H$) from
the best fit (Eq. \ref{lin}) linear/quadratic
relation.  Here we identified seven such outliers, removed them, and
refit the remaining 34 HII regions achieving the fit in
 (Eq. \ref{lin32}) which
has a much lower reduced chi-squared ($\chi ^2/dof = 0.96$).  In our
$Y$ vs. $O$ and $N$ fits we will use this modified, first cut$^\prime$
 set (34
regions = first cut set minus the seven outliers)
and compare with the first cut (41 regions) set.

In the previous section we have also considered an extremely metal-poor
subset of the data.  This second cut set consists of half of the
first cut set; 21 regions with $10^6 O/H \leq 80$.  Here, too, the
dispersion about the $N$ vs. $O$ fit is large ($\chi ^2/ dof = 1.5$)
and the two outliers from our first cut set with $10^6 O/H \leq 80$
are more than 2$\sigma$ discrepant here too.  Removing the two outliers
results in a modified, second cut$^\prime$ set (19
regions = second cut set minus the two outliers) with a much reduced chi-
squared ($\chi ^2 /dof = 0.93$) around the mean $N$ vs $O$ relation.
We have also fit the $Y$ vs. $O$ and $N$ data for these two sets.

In Fig. 4 we display the $Y$ vs. $O/H$ data for all 49 HII regions in
the PTSE and Skillman et al. data sets.
The seven regions eliminated by our first cut are shown as filled
triangles; note that the three highest $^4He$ abundances ($Y
\geq 0.26$) belong to these regions.  Also in Fig. 4 we have
distinguished with different symbols the C (open squares),
P (open circles) and D (open triangles) regions as well as
the seven outliers (filled circles) from the first cut set.

\vskip .5in
\begin{table}[h]
\centerline {\sc{\underline{Table 2:} Linear Fits for $Y$ vs. $O/H$}}

\vspace {0.1in}
\begin{center}
\begin{tabular}{|ccccccc|}                     \hline \hline
Set &  \# Regions & $r$ & ${\chi}^2/dof$ &  $Y_P$  &  $10^{-2}
 \times$ slope &$Y_P^{2\sigma}$ \\ \hline

All & 49 & 0.56 & 0.78 & $.234 \pm .003$ & $ 1.14 \pm 0.24$ & 0.239 \\
1st cut & 41 & 0.51 & 0.61 & $.232 \pm .003$ & $ 1.38 \pm 0.36$& 0.238 \\
-outliers & 34 & 0.45 & 0.70 &$.232 \pm .003$ & $1.39 \pm 0.38$& 0.238 \\
2nd cut & 21 & 0.41 & 0.64 & $.229 \pm .005$ & $2.37 \pm 1.13$ & 0.238 \\
-outliers & 19 & 0.40 & 0.70 & $.229 \pm .005$ & $2.42 \pm 1.15$ & 0.238\\
C & 22 & 0.35 & 0.71 & $.232 \pm .003$ & $1.58 \pm 0.54$ & 0.238 \\ \hline
\end{tabular}
\end{center}
\end{table}
\vskip .5in
\begin{table}[h]
\centerline {\sc{\underline{Table 3:} Linear Fits for $Y$ vs. $N/H$}}

\vspace {0.1in}
\begin{center}
\begin{tabular}{|ccccccc|}                     \hline \hline
Set &  \# Regions & $r$ & ${\chi}^2/dof$ &  $Y_P$
  &  $10^{-3} \times$ slope &$Y_P^{2\sigma}$ \\ \hline

All & 49 & 0.66 & 0.66 & $.236 \pm .002$ & $ 1.72 \pm 0.33$ & 0.240\\
1st cut & 41 & 0.57 & 0.58 & $.234 \pm .002$ & $ 2.71 \pm 0.68$  & 0.239 \\
-outliers & 34 & 0.48 & 0.69 &$.234 \pm .003$ & $2.77 \pm 0.76$ & 0.239\\
2nd cut & 21 & 0.47 & 0.63 & $.231 \pm .004$ & $4.85 \pm 2.27$ &0.239 \\
-outliers & 19 & 0.44 & 0.70 & $.232 \pm .004$ & $4.79 \pm 2.29$ & 0.239 \\
C & 22 & 0.46 & 0.60 & $.233 \pm .003$ & $3.62 \pm 1.17$ & 0.238\\ \hline
\end{tabular}
\end{center}
\end{table}

In Tables 2 and 3 we show the results of our linear least square
fits to $Y$ vs. $O/H$ and $Y$ vs. $N/H$ relations respectively.
We list the number of HII regions in each set we fit along with the
correlation coefficient (r), the reduced chi-squared of the fit ($\chi
^2/dof)$, the intercept (the inferred, zero-metallicity, primordial
$^4He$ abundance ($Y_P$)) along with its $1\sigma$ uncertainty, the slope
of the $Y$ vs. $O/H$ and $N/H$ relations respectively and,
in the last column, the $2\sigma$ (statistical) upper bound to
$Y_P$, ($Y_P^{2\sigma}$).

The linear fits in the tables are significantly better, statistically,
than is a simple weighted mean of the data. In addition, we have
tested three-parameter fits to the data and in these cases we have found
that the data is definitely not reliably correlated with respect to these fits.
Thus the linear fits, at present, offer the best representation of the data.
Notice that all the fits in tables 2 and 3 have very small reduced
chi-squareds and, that all the inferred primordial abundances ($Y_P$) in
these tables are mutually consistent.  From these results we may infer
that $Y_P = 0.232 \pm 0.003$ and $Y_P^{2\sigma} \leq 0.238$.  Notice, too,
the improvement in $\chi^2/dof$ between all (49) HII
regions and our first cut.  However, unlike the reduced dispersion in
the $N$ vs. $O$ relation, eliminating the outliers from the first cut or
second cut sets does not result in an improvement in $\chi^2/dof$.
And, if we compare the full first cut set with the C
set, it is unclear that eliminating regions with possible or detected
WR features, results in an improved fit with a reduced dispersion.

The results in Tables 2 and 3 confirm previous analyses (OSW; Pagel
et al. 1992) which found steep $Y$ vs. $O/H$ and $Y$ vs. $N/H$
relations.  For example, if $Z \approx 20 (O/H)$ then the $Y$ vs.
$O/H$ slopes in Table 2 correspond to $6 \lsim \Delta Y/\Delta Z  \lsim 12$.
Alternately if, for example, we evaluate the first cut fits at the solar
oxygen and nitrogen abundances respectively, we would predict
${Y_\odot}^{O/H} \approx 0.35$, ${Y_\odot}^{N/H} \approx 0.54$,
grossly in excess of the solar value $Y_\odot \approx 0.28$.  As
emphasized at the outset, there need be no connections between the
evolution of the extra-galactic HII regions and the solar vicinity of
the Galaxy.  The role of our $Y$ vs. $O$ and $Y$ vs. $N$ relations
inferred from the metal-poor extra galactic HII regions is simply to
aid in our extrapolations to the primordial abundance $Y_P$.  In
Figure 5 we show all the first cut data (outliers in filled symbols)
for $Y$ vs. $O/H$ along with the first cut fit from Table 2.  In Fig. 6
we show the corresponding data set for $Y$ vs. $N/H$
along with the first cut fit from
Table 3.

In OSW we explored an alternate approach to a 2$\sigma$ upper bound to
$Y_P$.  Consider the HII region with the lowest value of $Y+2\sigma$
(0.238 for I Zw18).  Since this - or any other of our set - region may
have been contaminated by stellar produced $^4He$, $Y_P^{2\sigma} \leq
(Y+2\sigma)_{min} = 0.238$, consistent with our $2\sigma$ bound from Tables 2
and 3.  However, as we average in the next lowest helium abundance
regions, although $\langle Y \rangle$ will increase, $\langle \sigma
\rangle$ will decrease $(\langle \sigma \rangle ^{-2} = {\sigma_1}^{-2} +
{\sigma_2}^{-2} + \dots)$ and, for some number of the HII regions,
$\langle Y \rangle + 2 \langle \sigma \rangle$ will achieve a
minimum (eventually, the increase in $\langle Y \rangle$
overwhelms the decrease in $\langle \sigma \rangle$) so that
$Y_P^{2\sigma} \leq (\langle Y \rangle + 2\langle \sigma \rangle )_{min}$.
For the PTSE and Skillman et al. data sets we find that $(\langle Y
\rangle + 2\langle \sigma \rangle )_{min} = 0.236$ so that $Y_P^{2\sigma}
\leq 0.236$.

Very recently, Skillman and Kennicutt (1993) and Skillman et al.(1994a)
have performed especially detailed and careful analyses of three of the
most metal-poor HII regions.  In order to assess the quality of their
derived statistical uncertainties and, to estimate some of the possible
systematic uncertainties, they have acquired data with several different
telescope/instrument combinations and they have taken great care in
reducing their data.  For each of these three HII regions (two in IZw18
and one in UGC4483) they derive helium abundances to better than 3\%
accuracy.  A weighted mean of their results provides an upper bound to
the primordial helium abundance: $Y_P\leq 0.234 \pm 0.004$ which is
competitive with those we have derived from some 3-4 dozen HII
regions.  This illustrates the potentially great value of very
detailed and careful analyses of a handful of the lowest metallicity
HII regions and we would urge observers to focus their efforts in
this direction.

In the above discussions we have, in preparation for our comparison
with the predicted BBN abundance $Y_{BBN}$, determined a $95\%$
CL upper bound to $Y_P$ based on the statistical uncertainties above:
$Y_P^{2\sigma}
\lsim 0.236-0.238$.  To have the most generous comparison, we will
adopt $Y_P^{2\sigma}
\leq 0.238$.  However, it must not be forgotten that there are
possible systematic uncertainties as well.  For example, although for
the high excitation metal-poor HII regions in our sample it is expected
that the HII and HeII zones coincide (Skillman et al. 1994a),
nonetheless there could
be differences at the 1-2\% level.  Similarly, although corrections for
collisional excitation (Ferland 1986; Clegg 1987) are estimated to be
negligible in most cases, 1-2\% corrections may not be excluded.  If
there is significant dust, not expected for our metal-poor HII regions,
then trapping of H-recombination photons followed by dust
absorption should be, but is generally not, accounted for (Baldwin et al. 1991;
Skillman \& Kennicutt 1993).  These, and
possibly other systematic effects, suggest that a one sigma estimate of the
systematic uncertainty is $\sigma_{syst} \approx 0.005$.  Thus, in
our comparisons discussed next, we will use $Y_P^{2\sigma}
\leq 0.238$ and $Y_P^{2\sigma} + \sigma_{syst}
\leq 0.243$ as our estimates for
$Y_P^{MAX}$.

\section{Discussion}

The fits to all the data sets in Tables 2 and 3 are mutually consistent
with a zero-metallicity, primordial $^4He$ mass fraction
\beq
Y_P = 0.232 \pm 0.003 \pm 0.005
\label{yp}
\eeq
Notice that for the full (49 regions) data set, which extends to higher
metallicity, the slopes in the $Y$ vs. $O/H$ and $Y$ vs. $N/H$
relations are shallower and, the intercepts, $Y_P$ correspondingly
higher $Y_P^{all} \approx 0.235 \pm 0.003$.  Nonetheless, all the fits
are consistent with a
two-sigma (statistical) upper bound of
\beq
Y_P^{2\sigma} \leq 0.238
\label{stat}
\eeq
To account for possible systematic uncertainties we have adopted a $\sim
2\%$ estimate, $\sigma_{syst}\approx 0.005$.
Thus, in our comparisons with the predictions of BBN, we shall use
the statistical upper bound in (\ref{stat}) as well as a maximum primordial
abundance of
\beq
Y_P^{MAX} = Y_P^{2\sigma} + \sigma_{syst} \leq 0.243
\label{max}
\eeq
We will also consider the uncertainties in our results from
uncertainties in our adopted values of $Y_P^{2\sigma}$ and $Y_P^{MAX}$.

In his recent thesis, Kernan (1993) has considered in great detail the
ingredients necessary for an accurate calculation of $Y_{BBN}$.
The work of Kernan (1993), Seckel (1994) and Gyuk and Turner
(1994) has led to small but significant corrections to the calculation
of $Y_{BBN}$ in WSSOK.  These differences have been summarized by
Kernan, Steigman and Walker (1994) who find for the standard case
of $N_\nu = 3$ and for the same adopted neutron lifetimes ($\tau_n$)
\beq
Y_{BBN}(K) - Y_{BBN}(WSSOK) = 0.0021 +0.0004\ln \eta_{10}
\eeq
Thus, in the ``interesting" range of nucleon to photon ratio
 $2~\lsim \eta_{10} \lsim 4, Y_{BBN}(K) - Y_{BBN}(WSSOK)
 = 0.0024-0.0027$.  Furthermore, since WSSOK, the $2\sigma$ lower bound to
the neutron lifetime has increased (Review of Particle Properties 1992)
from $\tau_n \ge
882 {\rm s}$ to $\tau_n \ge 885{\rm s}$ and this adds 0.0006
 to the WSSOK results.  Thus, overall, the
predicted $Y_{BBN}$ has increased by $\approx 0.003$ at fixed
$\eta_{10}$; with the {\em same} observational upper bounds to $Y_P$
(OSW), this would result in reduced upper bounds to $\eta_{10}$ and
$N_\nu$.  The constraints we present here are based on the BBN
calculations of Kernan (1993) and Kernan, Steigman and Walker
(1994) and the upper bounds to $Y_P$ in equations (\ref{stat}) and (\ref{max}).

First let us consider the upper bound to the nucleon abundance,
$\eta_{10}$ which follows from the upper bound to $Y_P$ and from
$Y_{BBN}$ with $N_\nu = 3$ and $\tau_n \ge 885{\rm s}$. For
$Y_P \le 0.238 (0.243)$,
\beq
\eta_{10} \le 2.5 (3.9)
\label{eta}
\eeq
In the past, (e.g. in WSSOK) the logarithmic dependence of $Y_{BBN}$
on $\eta$  has prevented us from using $Y_P$ to provide a
significant bound to $\eta$.  This effect is still noticeable in (\ref{eta}).
Nonetheless, even our conservative bound $Y_P^{MAX} \le 0.243$, combined with
the newer calculations of $Y_{BBN}$, does lead to a restrictive upper
bound (e.g. in WSSOK it was the primordial abundance of $^7Li$
which was used to provide the bound $\eta_{10} \le 4.0$).  And, the
more restrictive statistical bound, $Y_P^{2\sigma} \le 0.238$, leads to an
upper bound to $\eta$ in apparent conflict with the lower bound
of $\eta_{10} > 2.8$ from
WSSOK.  Before exploring the predicted lower bounds to the
primordial abundances of $D$ and $^3He$ from the upper bounds to
$\eta$ in (\ref{eta}), we note that the uncertainty in $\eta$ is related to
the uncertainty in $Y_P$ by
\beq
{\Delta \eta \over \eta} \approx {\Delta Y \over 0.012}
\eeq
which for $\Delta Y = 0.001$ and $\eta_{10} = 2.5(3.9)$, corresponds to
$\Delta \eta \approx 0.21 (0.33)$.

The predicted primordial abundances of $D$ and of $^3He$ decrease
with increasing $\eta_{10}$.  For the upper bounds to $\eta_{10}$ in
(\ref{eta}), standard BBN calculations yield (WSSOK; Kernan (1993)),
\begin{eqnarray}
10^5(D/H)_P \ge 10.1 (4.9) \\
10^5(^3He/H)_P \ge 1.7 (1.4)
\end{eqnarray}
The abundances of $D$ and $^3He$ for $\eta_{10} \leq 2.5$ ($Y_P \le
0.238$) are large and possibly in conflict with the solar system and
interstellar data (WSSOK; Steigman and Tosi 1992); however, see
Vangioni-Flam, Olive, and  Prantzos (1994). The problem here, is that
today the ISM deuterium abundance is observed to be $10^5 (D/H) = 1.5$
(Linsky et al. 1992) with very small errors. This would require a
destruction factor of nearly 7. While models were
 found (Vangioni-Flam et al. 1994)
which could
destroy deuterium by a factor of 5 (the value needed when $\eta_{10} = 3$)
and could probably be pushed to get the additional deuterium destruction,
the real problem lies with $^3He$.
Models which destroy deuterium tend to produce $^3He$ and yield too
large a value for the sum $(D +~^3He)/H$ when evaluated at the age
corresponding
to the formation of the solar system.  Unless stellar models for the survival
of $^3He$ in low mass stars have been overestimated, this constraint will
be difficult to overcome.

 For $\eta_{10} \le 3.9$
($Y_P \leq 0.243$) there is consistency with earlier analyses (e.g.,
WSSOK) which now require $2.8 \leq \eta_{10 }\leq 3.9$.
Corresponding to any uncertainties in our adopted upper bounds to
$Y_P$, there are uncertainties in the predicted lower bounds to
primordial $D$ and $^3He$,
\begin{eqnarray}
\Delta y_{2P}/y_{2P} \approx  -\Delta Y_P/0.007 \\
\Delta y_{3P}/y_{3P} \approx  -\Delta Y_P/0.024
\end{eqnarray}
So, for $\Delta Y_P \approx \pm 0.001$ and $10^5y_{2P} = 10.1
(4.9), 10^5 \Delta y_{2P} \approx \pm 1.4 (0.7)$; for
$\Delta Y_P \approx \pm 0.001$ and $10^5y_{3P} = 1.7
(1.4), 10^5 \Delta y_{3P} < \pm 0.1$.
The bottom line is that, if $Y_P \le 0.238$, the predicted primordial
abundance of D is large and potentially detectable (Webb et al. 1991;
Carswell et al. 1994;
Songaila et al. 1994) in the QSO
Lyman- $\alpha$ absorption systems.  This will provide a crucial test of the
consistency of BBN and the standard hot big bang cosmology.

Upper bounds to $\eta_{10}$ imply upper bounds to the cosmological
density of baryons (nucleons), with implications for the question of
the nature of the cosmologically dominant dark matter.  Using
$\Omega_{BBN}$ for the baryon density parameter from BBN,
a CBR temperature of $2.726\pm0.010K$ and, a Hubble parameter
of $H_o = 50 h_{50} {\rm kms}^{-1}{\rm Mpc}^{-1}$,
for $\eta_{10} \le 2.5 (3.9)$, we predict
\beq
\Omega_{BBN} h_{50}^2 \le 0.037 (0.058)
\eeq
For a lower bound to the Hubble parameter of $H_0 \ge 40 (h_{50}
\ge 0.8)$ we find an upper bound to $\Omega_{BBN} \le
0.058 (0.090)$ which makes the case for non-baryonic
dark matter even stronger.

Finally we turn to the implications for particle physics, specifically
physics beyond the standard model, of our upper bounds to
primordial helium.  The above comparisons have been for the
standard case of three massless (or, very light: $\lsim 0.1$ MeV)
neutrino flavors, $N_\nu = 3$.  The effect of the presence of ``new"
particles (beyond the standard model) which enhance the total
energy density at BBN is to speed-up the universal expansion rate
(Schvartzman 1969) and leave available more neutrons to form more
$^4He$.  Thus, the predicted helium mass fraction increases with $
N_\nu \ge 3$ (Steigman, Schramm and Gunn 1977): $\Delta Y_{BBN}
\approx 0.012 \Delta N_\nu$.  Now, recall that compared to WSSOK
the predicted $Y_{BBN}$ - for the same $N_\nu$ - has increased by $\sim
0.003$.  Thus, our current comparison, even if we used the WSSOK
value of $Y_P \le 0.240$, will yield a much more restricted upper
bound to $N_\nu$.
In the past, to find the new upper bound to $N_\nu$, we compared the
upper bound to $Y_P$ with
the lower bound to $Y_{BBN}$ evaluated for $\tau_n \ge 885$ s
and $\eta \ge \eta_{min}$, the most favorable limits -in the
sense that they maximize $N_\nu$- for these quantities.  In
WSSOK the bound to the primordial abundance of $D$ + $^3He$,
inferred from solar system and interstellar observations, was used to
bound $\eta_{10} \ge 2.8$.  We have already noted above that the
2$\sigma$ statistical upper bound, $Y_P^{2\sigma} \le 0.238$,
 corresponds to $\eta_{10} \le 2.5$ so
that, for this bound to $Y_P$, we will find $N_\nu < 3$.  On the other hand,
including an estimate of the possible systematic uncertainty to bound
$Y_P^{MAX} \le 0.243$, permits  $N_\nu > 3$ but leads to a slightly more
restrictive bound than that in WSSOK.  Thus for $Y_P \le 0.238 (0.243)$
we now find
\beq
N_\nu \le 2.9 (3.3)
\label{nnu}
\eeq
This bound, however, is not really a 2$\sigma$ upper bound to $N_\nu$ as
each of $Y_P$, $ \eta_{10}$, and $\tau_n$ have been allowed to take their
extreme
values. If we take for the observed value, $Y_P =  0.232 \pm 0.003
\pm 0.005$, and we use the BBN prediction of $^4He$
evaluated for $\tau_n = 889.1 \pm 2.1$~s and
$\eta_{10} = 3.0 \pm 0.3$,
 we can derive the best fit value of $N_\nu$. Note that the latter
value for $\eta$ is chosen for consistency with the other
light elements $D$, $^3He$, and $^7Li$.
The best fit value of $N_\nu$ now becomes
\beq
N_\nu = 2.17 \pm 0.27 \pm 0.42
\eeq
Thus a $2\sigma$ upper limit (statistical) would be $N_\nu < 2.71$ and
$N_\nu = 3$ is consistent at the 3.1$\sigma$ level.  When ``systematics" are
included, however, we see that the 2$\sigma + \sigma_{syst}$ upper bound to
$N_\nu$ is 3.13 which is consistent with standard model physics.
Note that the bounds in (\ref{nnu}) correspond to the upper bound on $D$ +
$^3He$,
$10^5 y_{23P} \le 10.0$.  Corresponding to any uncertainty in this
bound, there will be uncertainty in $N_\nu^{MAX}$.  For example, if the
$D$ + $^3He$ bound is only relaxed from $10^5y_{23P} \lsim 10.0$ to  $\lsim
11.7$, $Y_P \lsim 0.238$ is then consistent with $N_\nu \lsim 3.0$.
Alternatively, the possible inconsistency in (\ref{nnu}), $N_\nu^{BBN} < 3$,
could be evidence for a massive ($\rsim $5-10 MeV), unstable ($\lsim $40 sec.)
tau-neutrino (Kawasaki et al. 1994).

\section{Summary}

The availability of a large data set (PTSE) of homogeneously analyzed
observations of metal-poor extra-galactic HII regions, recently
supplemented by a sample of very metal-poor regions (Skillman et al.
1994b) encouraged us to employ this data to derive the primordial
abundance of $^4He$ which is key to testing the consistency of Big Bang
Nucleosynthesis.  Our goal has been to define a mean relation
between the observed abundances of oxygen and/or nitrogen and
helium and, to use this relation as an aid in extrapolating to the
zero-metallicity, primordial helium abundance.  As an aid in establishing
this relation, we first analyzed the nitrogen and oxygen data to see if
we could identify regions with excess or depleted nitrogen or oxygen
which might contribute to dispersion around a mean helium vs.
nitrogen or helium vs. oxygen relation.  For our ``first cut" data set,
with $10^6(O/H) \le 150$ and $10^7(N/H) \le 100$,
 we found that the data is reasonably well fit
by either a power law ($N/H \sim (O/H)^\alpha$, with $1 \lsim \alpha  \lsim 2$
or a
linear/quadratic relation ($N/H \sim a(O/H) + b(O/H)^2$).  In the latter case,
the
linear (``primary") term dominates over the quadratic (``secondary")
term at low metallicity (typically, for $O/H \lsim 10^{-4}$).  We used the
latter fit to identify ``outliers", HII regions which differ by more than
$2\sigma$ from the mean $N$ vs. $O$ relation.  With such
discrepant regions removed the reduced chi-squareds about the
mean relations are quite small.  We have emphasized that the metal-
poor extra-galactic HII regions are likely a very heterogeneous set
which need not give us a glimpse of the early evolution of nitrogen
and oxygen in the Galaxy.  Nonetheless, our simple power-law and
linear/quadratic relations may be extrapolated to provide not bad fits to the
observed nitrogen and oxygen abundances in halo
stars, Galactic HII regions and, even the sun.

Next, we fit the observed helium mass fractions to linear relations
with $O/H$ and $N/H$.  For all of our subsets of the data (based on
$N$ vs. $O$) we find mutually consistent fits with small dispersion around the
mean relations and we infer:  $Y_P = 0.232 \pm 0.003$, with a two
$\sigma$ (statistical) upper bound $Y_P^{2\sigma} \le 0.238$.  We
confirm previous analyses which found steep slopes in the $Y$ vs.
$O$ and $Y$ vs. $N$ relations ($\Delta Y/\Delta Z \approx 6-12$).
  Thus, it is
clear that these relations cannot be extrapolated to interstellar or
solar metallicities.

It is much more difficult to estimate the magnitude of possible
systematic uncertainties.  Here, we have agreed with previous
analyses and adopted a $2\%$ uncertainty which corresponds to
$\sigma_{syst} \approx 0.005$.  This then led us to a ``maximum"
value for primordial helium:  $Y_P^{MAX} = Y_P^{2\sigma} + \sigma_{syst}
\le 0.243$.  We have used both bounds $Y_P^{2\sigma}$ and
$Y_P^{MAX}$, in our comparisons with the predictions of BBN.

Recent attempts (Kernan 1993; Seckel 1993; Gyuk and Turner 1993)
to calculate to high accuracy the BBN yield of $^4He$ have led to
an overall increase in $Y_{BBN}$ ($\sim 0.003$ compared to WSSOK;
residual uncertainty, including the uncertainty in the neutron
lifetime, $\lsim 0.002)$. Thus, the comparison between $Y_P$ and
$Y_{BBN}$ leads to tighter constraints on the nucleon-to-photon ratio
($\eta_{10}$) and on the number of equivalent light neutrino flavors
($N_\nu$).  For $Y_P \leq Y_P^{2\sigma} \le 0.238$, we found $\eta_{10}
\leq 2.5$ and $N_\nu \leq 2.7$.  This suggestion of a crisis in BBN
does, however, depend crucially on the precise value of the
primordial abundance of $D +~^3He$.  There is apparent inconsistency
(over-production of $D$ and $^3He$) if $10^5y_{23P} \le 10.0$
(WSSOK), but, there would be consistency for the weaker bound of
$10^5y_{23P} \le 11.7$.  If, indeed, the primordial abundance of $D$
were as large as $10^5y_{2P} \approx 10$, deuterium may well be
observable in the QSO Lyman - $\alpha$ absorption systems.

Our bounds to the nucleon abundance ($\eta_{10} \le 2.5(3.9)$)
imply reduced upper bounds to the nucleon density in the Universe:
$\Omega_{BBN}h_{50}^2 \lsim 0.04 (0.06);$ for $h_{50} \ge 0.8,
\Omega_{BBN} \lsim 0.06$ (0.09).  These lower upper
bounds (compared, e.g., to WSSOK) strengthen the case for non-
baryonic dark matter.

The 2$\sigma$ statistical upper bound to $Y_P$ is sufficiently small
that, if the primordial abundance of $D$ + $^3He$ is no larger than
$10^5y_{23P} = 10$, then the bound on the number of equivalent, light
($\le 10MeV$) neutrinos is $N_\nu \le 2.7$, in apparent conflict
with the LEP result, ($N_\nu^{LEP} = 3$).  However, recall that LEP will
probe neutrinos as massive as $M_Z/2$ while BBN is sensitive
to light neutrinos.  Indeed, Kawasaki et al. (1994) have noted that a
massive ($5-10 \lsim m_\nu \lsim 31 MeV$), unstable ($\tau_\nu \lsim 40$
sec) tau-neutrino would lead to lower $^4He$ production than in the
case of three massless neutrinos.

On the other hand, with allowance for possible systematic
uncertainties, $N_\nu \leq 3.1$ is allowed for $10^5y_{23P} \leq 10$.
Thus, there may be no conflict with the standard model and, very
little room for new light particles beyond the standard model.

\vskip 1.0truecm
\noindent {\bf Acknowledgements}
\vskip 1.0truecm
We would like to thank E. Skillman, R.J. Terlevich, R.C. Kennicutt,
D.R. Garnett, and E. Terlevich for sharing their prepublished data.
We would also like to thank B. Pagel, E. Skillman, and T.P. Walker
 for useful discussions.  This paper has been completed while GS
is an Overseas Fellow at Churchill College and a Visiting Scientist
at the Institute of Astronomy (Cambridge, England) and he thanks
them for hospitality.
His work was supported in part by  DOE grant DE-FG02-94ER-40823.
The work of KAO was in addition supported by a Presidential Young
Investigator Award.
\newpage
\beginapjbib

\bibitem Balbes, M.J. Boyd, R.N. \& Mathews, G.J. 1993, ApJ, 418, 229

\bibitem Baldwin, J.A., Ferland, G.J., Martin, P.G., Corbin, M.R.,
Cota, S.A., Peterson, B.M., \& Slettebak, A. 1991, ApJ, 374, 580

\bibitem Barbuy, B.\& Erdelyi-Mendes, M. 1989 Astron. \& Astophys., 214, 239

\bibitem Bevington, P.R. 1969 Data Reduction and Error
 Analysis for the Physical Sciences (New York: McGraw Hill)

\bibitem Boesgaard, A. \& Steigman, G. 1985, Ann. Rev.  Astron. Astrophys.,
23,  319

\bibitem Brocklehurst, M. 1972, MNRAS, 157, 211

\bibitem Campbell, A. 1992, ApJ, 401, 157

\bibitem Carswell, R.F., Rauch, M., Weymann, R.J., Cooke, A.J.,
\& Webb, J.K. 1994, MNRAS (in press)

\bibitem Clegg, R.E.S. 1987, MNRAS, 229, 31P

\bibitem Cunha, K. \& Lambert, D.L. 1992, ApJ, 399, 586

\bibitem De Young, D.S., \& Gallagher, J.S. 1990, ApJ (Lett.) 356, L15

\bibitem Ferland, G.J. 1986, ApJ, 310, L67

\bibitem Fuller, G.M., Boyd, R.N., \& Kalen, J.D. 1991, ApJ, 371, L11

\bibitem Gies, D.R. \& Lambert, D.L. 1992, ApJ, 387, 673

\bibitem Grevesse, N. \& Anders, E. 1989 in {\sl Cosmic Abundances of Matter},
ed. C.J. Waddington (AIP Conf. Proc. 183), p. 1.

\bibitem Gyuk, G.  \& Turner, M.S. 1994, preprint

\bibitem Kawasaki, M., Kernan, P., Kang, H.-S., Scherrer, R.J.,
Steigman, G., \& Walker, T.P. 1994, Nucl. Phys. B,
(in press, vol.419)

\bibitem Kernan, P. 1993, OSU Physics Ph.D. Thesis

\bibitem Kernan, P., Steigman, G. \& Walker, T.P. 1994, (In Preparation)

\bibitem Kunth, D. \& Sargent, W. 1983, ApJ, 273, 81

\bibitem Lequeux, J., et. al. 1979, A \& A, 80, 155

\bibitem Linsky, J.L., et al. 1992, ApJ, 402, 694

\bibitem Mathew, G.J., Boyd, R.N., \& Fuller, G.M. 1993, ApJ, 403, 65

\bibitem Olive, K.A., Steigman, G. \& Walker, T.P. 1991, ApJ, 380, L1

\bibitem Pagel, B.E.J. 1993, Proc. Nat. Acad. Sci., 90, 4789

\bibitem Pagel, B.E.J. \& Kazlauskis, A. 1992, MNRAS, 256, 49P

\bibitem Pagel, B E.J., Simonson, E.A., Terlevich, R.J.
\& Edmunds, M. 1992, MNRAS, 255, 325 (PSTE)

\bibitem  Pagel, B.E.J., Terlevich, R.J. \& Melnick, J. 1986,
 PASP, 98, 1005 (PTM)

\bibitem Peimbert, M. \& Torres-Peimbert, S. 1974 Ap J, 193, 327

\bibitem Reeves, H. 1993, Rev. Mod. Phys. (In Press)

\bibitem Review of Particle Properties 1992, Phys. Rev. D45, 1

\bibitem Searle, L. \& Sargent, W.L.W. 1971, ApJ, 173, 25

\bibitem Seckel, D. 1994, Preprint

\bibitem Schvartzman, V. 1969, JETPL, 9, 184

\bibitem Skillman, E., \& Kennicutt 1993, ApJ, 411, 655

\bibitem Skillman, E., Terlevich, R.J., Kennicutt, R.C., Garnett, D.R.,
\& Terlevich, E. 1994a, ApJ, (in press)

\bibitem Skillman, E., et al. 1994b, ApJ Lett (in preparation)

\bibitem Smits, D.P. 1991, MNRAS, 251, 316

\bibitem Sneden, C., Lambert, D.L. \& Whitaker, R.W. 1979 ApJ, 234, 964

\bibitem Songaila, A., Cowie, L.L., Hogan, C. \& Rugers, M.,
1994, Nature, 368, 599

\bibitem Steigman, G., Schramm, D.N. \& Gunn, J. 1977, Phys. Lett., B66, 202

\bibitem Steigman, G. \& Tosi, M. 1992, ApJ, 401, 150

\bibitem Vangioni-Flam, E., Olive, K.A., \& Prantzos, N. 1994 ApJ,
(in press)

\bibitem Walker, T. P., Steigman, G., Schramm, D. N., Olive, K. A.,
\& Kang, H. 1991 ApJ, 376, 51 (WSSOK)

\bibitem Webb, J.K., Carswell, R.F., Irwin, M.J. \& Penston, M.V., 1991,
MNRAS, 250, 657

\bibitem Wheeler, J.C., Sneden, C.\& Truran, J.W. 1989 Ann. Rev. Astron. \&
Astrophys, 27, 279

\endapjbib

\newpage
\noindent{\bf{Figure Captions}}

\vskip.3truein

\begin{itemize}

\item[]
\begin{enumerate}
\item[]
\begin{enumerate}
\item[{\bf Figure 1:}] The nitrogen and oxygen abundances observed for all 48
HII regions
 in the  PSTE  and Skillman et al. (1994b) data sets. Our ``first cut" data set
is indicated by the dashed lines and our ``second cut" data set is further
restricted by the dotted line.

\item[{\bf Figure 2:}] $N/O$ plotted vs. $O/H$ for the first cut sample of
40 HII regions. Also shown is the linear fit (Eq. \ref{lin}) derived for this
data.

\item[{\bf  Figure 3:}] The same data as in Figure 2 is shown and the 2$\sigma$
outliers are identifed by filled circles.  Also shown is the new linear fit
with the outliers removed from the data set.

\item[{\bf  Figure 4:}]  The helium and oxygen abundances for all 48 HII
regions. The eight regions eliminated by our first cut are shown as filled
triangles.  Also  shown are
the eight outliers (filled circles) from the first cut set.
The remaining points are distinguished by their observed WR
features:  C (open squares),
P (open circles) and D (open triangles).

\item[{\bf  Figure 5:}]   $Y$ vs. $O/H$ for the first cut data set of 40 HII
regions. Outliers in $N/O$ vs. $O/H$ are shown as filled circles.

\item[{\bf  Figure 6:}]   $Y$ vs. $N/H$ for the first cut data set of 40 HII
regions. Outliers in $N/O$ vs. $O/H$ are shown as filled circles.

\end{enumerate}
\end{enumerate}
\end{itemize}

\end{document}